\documentclass[11pt]{article}
\usepackage{authblk}
\usepackage[a4paper,textwidth=30pc,textheight=46pc]{geometry}

\usepackage{float}
\usepackage{graphicx}
\usepackage{url}

\begin{document}

\title{Functional Percolation:\\ Criticality of Form and Function}

\author{Galen J. Wilkerson}
\affil{Komplexity AI LLC}
\affil{Correspondence: \texttt{gwilkerson@komplexai.io}}
\affil{ORCID: 0000-0002-7957-5821}

\date{\today}

\maketitle

\begin{center}
Published in \textit{Journal of Complex Networks} (2026).\\
DOI: \url{https://doi.org/10.1093/comnet/cnag030}
\end{center}

\begin{abstract}

Understanding how network structure constrains and enables information processing is a central problem in statistical mechanics of interacting systems \cite{Broadbent1957,Stauffer1994Percolation}. Here we study random networks across the structural percolation transition and analyze how connectivity governs realizable input--output transformations under cascade dynamics. Using Erd\H{o}s--R\'enyi networks as a minimal ensemble, we examine structural, functional, and information-theoretic observables as functions of mean degree. We find that the emergence of the giant connected component coincides with a sharp transition in realizable information processing: complex input--output response functions become accessible, functional diversity increases rapidly, output entropy rises, and directed information flow---quantified by transfer entropy---extends beyond local neighborhoods. We term this coincidence of structural, functional, and informational transitions \emph{functional percolation}, referring to a sharp expansion of the space of realizable input--output functions at the percolation threshold. Near criticality, networks exhibit a Pareto-optimal tradeoff between functional complexity and diversity, suggesting that percolation criticality may provide a general organizing principle of information processing capacity in systems with local interactions and propagating influences.
\end{abstract}


\medskip
\noindent\textbf{Code and data availability:} Code and supplementary materials are available at \url{https://github.com/galenwilkerson/Functional-Percolation}.

\section{Introduction}

\begin{figure}[H]
    \vspace*{-0.75\baselineskip}
    \centering
    \includegraphics[width=\linewidth]{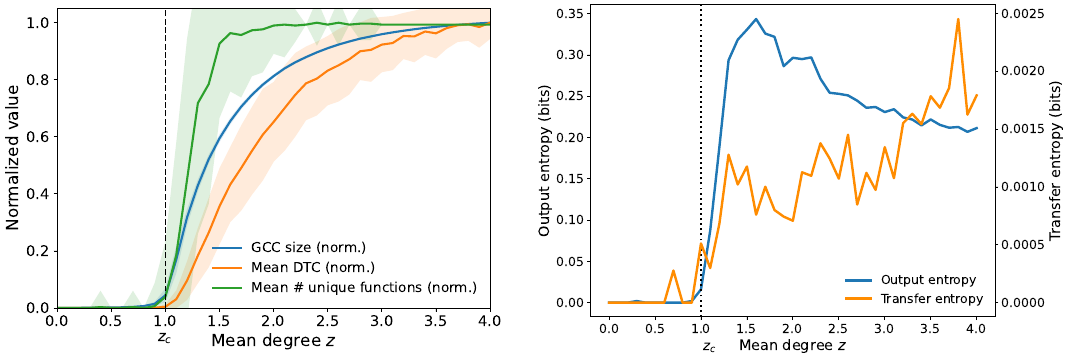}
\caption{
\textbf{Structural, functional, and information-theoretic signatures of percolation in random cascade networks.}
\textit{Left:} Mean structural and Boolean-functional order parameters as a function of mean degree $z$, shown in normalized units. The emergence of the giant connected component (GCC) coincides with a sharp onset of typical decision-tree complexity (DTC) and a rapid expansion in the mean number of unique Boolean response functions realized by the network, indicating a transition in realizable functional capacity and diversity at the structural percolation threshold $z_c$. Shaded regions indicate $\pm 1$ standard deviation across network realizations, highlighting enhanced fluctuations near criticality.
\textit{Right:} Information-theoretic measures versus connectivity. Output entropy (left axis) and transfer entropy (right axis) both increase sharply near $z_c$, reflecting the onset of expressive response diversity and directed information flow once global connectivity is established. Output entropy exhibits a peak near criticality, while transfer entropy continues to increase with connectivity. 
}
    \label{fig:functional_percolation}
    \par\smallskip\noindent{\footnotesize\textit{Alt text:} Two line plots against mean degree $z$, each marked with a vertical line at the critical value $z_c$ near $1$. Left panel: three normalized curves---giant connected component size, mean decision-tree complexity, and mean number of unique Boolean functions---all near zero below $z_c$ and rising sharply to a plateau above it, with shaded one-standard-deviation bands widest near $z_c$. Right panel: output entropy rises steeply at $z_c$ to a peak just above it and then declines slightly, while transfer entropy (right axis) rises at $z_c$ and keeps increasing with mean degree.}
\end{figure}

It has been widely recognized that network cascades exhibit hallmarks of universal spreading processes, closely tied to percolation \cite{bak1988self,sethna2001crackling,watts2002simple}.  
The universality of network cascades appears to arise from the fundamental nature of their minimal physical ingredients:  State encoding by system elements, thresholding, information propagation, and energy input to drive the cascade process.  These qualities can be found in many neuronal, social, biological, and physical systems.  
One may also reasonably conjecture that cascades form a general physical substrate for information processing, as node response patterns to inputs constitute Boolean logic functions \cite{Wilkerson2022}.   
Arguably, due to this dependence on very simple physical constraints and mechanisms, cascades may represent a particularly minimal and broadly applicable class of Boolean network dynamics \cite{Kauffman1969,watts2002simple}.
While microscopic interactions are continuous, coarse-graining generically produces state-dependent, threshold-like transitions, making cascade dynamics a widely applicable effective description of information propagation in interacting systems.

As the highest entropy, least structured systems, \textit{random networks} undergoing cascades provide a minimally-constrained physical basis for information processing, and therefore give an upper bound on structural degrees of freedom available to any realizable system.  Thus, studying the emergence of worst-case information processing establishes a broadly applicable baseline of what the heterogeneous systems found in nature can achieve \cite{newman2018networks}.

One may also argue for the near-universality of Boolean functions, as they simply characterize patterns of coarse-grained responses to input stimuli, and do not only concern computers, but more fundamentally the simplest units of information \cite{crutchfield1994calculi,shannon1948mathematical}.
These functions are also requisite for any system that performs learning or information processing under some level of coarse-graining or decision-making \cite{minsky2017perceptrons,shannon1948mathematical}.

Together, these points suggest that cascades and their induced Boolean transformations are generic computational responses in a large class of systems with local state, thresholds, and propagating interactions. They therefore provide a natural baseline for understanding how complex information processing emerges in physical and biological systems.
Although we have not formalized an axiomatic framework here, the minimality and ubiquity of these ingredients suggest that cascades are promising candidates for developing a possible principled axiomatization of information processing in physical systems.

Threshold cascade conditions on networks have been studied extensively, including exact results on seed-size--dependent global cascades \cite{gleeson2007seed}. While these works characterize when cascades occur, functional percolation concerns the accessibility and diversity of global response functions enabled by cascades, a distinct question not addressed in prior cascade-percolation analyses

\medskip
\noindent
\textbf{In this work, we make this intuition explicit.}
We show that while percolation in random networks enables large-scale information integration through the emergence of the giant connected component, the resulting Boolean response space grows combinatorially faster than the underlying configuration space of the system.  This scale mismatch poses a fundamental challenge for information processing in physical networks, demanding that systems exploit all available dynamical and structural mechanisms.  We demonstrate that criticality at the percolation transition provides a natural resolution of this challenge, simultaneously maximizing functional complexity, diversity, and information flow within this cascade model, a phenomenon we term \emph{functional percolation}.

To our knowledge, this work provides systematic evidence that structural percolation induces a sharp transition in the space of realizable Boolean functions and associated information-theoretic observables.

\section{The Core Problem: Scale Mismatch Between Form and Function}

\begin{figure}[h!]
    \centering
    \includegraphics[width=\linewidth]{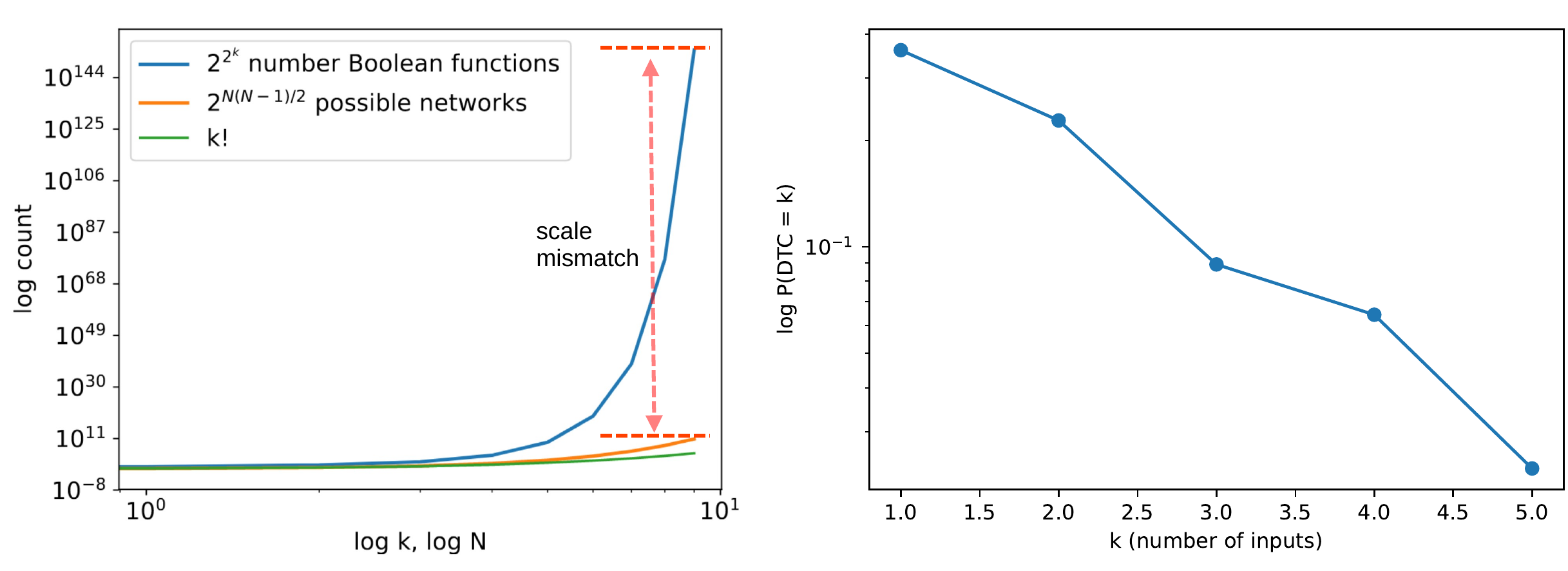}
   \caption{\textbf{Scale mismatch of form and function.}  [Left] The number of configurations of a system of $N$ elements is exponentially overwhelmed by the number of possible response patterns (Boolean functions) to input perturbations $k$.  [Right]  The probability of a random network to achieve complex functions decreases exponentially in $k$.  } 
      \label{fig:scale_mismatch}
    \par\smallskip\noindent{\footnotesize\textit{Alt text:} Two plots. Left panel: on a logarithmic count axis versus $\log k$ and $\log N$, the number of Boolean functions ($2^{2^k}$) rises far more steeply than the number of possible networks ($2^{N(N-1)/2}$) and $k!$; a dashed arrow marks the large ``scale mismatch'' gap between the Boolean-function curve and the network curve at large argument. Right panel: the probability that a random network realizes a fully $k$-dependent function, on a logarithmic axis, decreases steadily as the number of inputs $k$ increases from $1$ to $5$.}
\end{figure}

Any system depending on these cascades for information processing is immediately faced by a challenge, when we pursue the above line of reasoning:  The number of possible Boolean functions (response patterns) dominates the number of configurations of any system implementing them.
For randomly connected systems, the number of possible configurations of $N$ elements grows as $2^{N(N-1)/2}$.  Meanwhile, the abstract number of possible functions on $k$ input stimuli grows as $2^{2^k}$, vastly outpacing the configuration space [Figure \ref{fig:scale_mismatch}].

Cascades in random networks face an additional challenge for information processing.  Functions that integrate information from many inputs, having high decision-tree complexity (DTC), are inherently improbable to achieve, as they require connected paths to many input nodes.  Such high-DTC functions are exponentially improbable in random networks \cite{Wilkerson2022}, despite the fact that they dominate the abstract function space, making them \textit{doubly} difficult to explore.

This puts a finger on the problem:  It is exactly these high DTC (having large $k$ inputs) functions which are most useful, as they represent integration of high-dimensional information, but is also exactly the space of these high-DTC functions which grows very large and hard to explore.
This reveals a \textit{scale mismatch between form and function}.

Thus, finite physical networks are faced with the overwhelming challenge of exploring the vast Boolean function space.  While it may be possible to explore this space by increasing their size, this would quickly lead to gross inefficiencies and become physically untenable.  The number of system elements would have to scale super-exponentially with the number of input stimuli, for networks to explore the logic circuits computing all possible Boolean functions.   This strongly suggests that \textit{networks must use every possible tool at their disposal to efficiently explore the function space}.

\section{Criticality and Functional Percolation}

In the universal information processing by cascades in random networks, outlined above, the ability of such systems to perform integrative information processing depends on large-scale connectivity \cite{Wilkerson2022}.  Integrative responses are not possible in fragmented random systems.  The emergence of global connectivity comes in the form of structural percolation, with the emergence of the giant connected component (GCC) at the critical connectivity $z_c$ \cite{Erdos1960RandomGraphs,Stauffer1994Percolation}.  Thus the GCC size versus connectivity is a macroscopic order parameter \cite{Stanley1971}.

Via cascades on the underlying emergent GCC, this \textit{symmetry breaking of form maps to a symmetry-breaking of function}.  The network topology, by realizing computational graphs, constrains the realizable functional responses.  That is, there is a difference between the double-exponential \textit{abstract} function space on $k$ inputs, and the exponentially scaling \textit{physically accessible function space}.  In this sense, functional percolation represents not merely the emergence of connectivity, but a phase transition-like change in the realizable information processing of physical systems, arising due to the coincidence of structural and functional transitions.

One may observe and measure this functional, information-processing phase transition of networks by several order parameters. In the Boolean function space, the mean decision-tree complexity (DTC) across realized functions captures the network's typical level of \textit{integrative processing}, while fluctuations across realizations provide a susceptibility-like measure of functional criticality, analogous to enhanced response fluctuations near phase transitions.  The number of unique Boolean functions realized quantifies \textit{functional diversity}.

Two entropic order parameters may also be observed to undergo a phase transition at criticality. The output entropy of non-input nodes characterizes the diversity and balance of possible responses --- that is, the \textit{expressivity} of the system under uniform input stimulation.  Transfer entropy (TE) characterizes directed \textit{information flow} through the network, reflecting dynamical responses to inputs in time \cite{shannon1948mathematical,schreiber2000measuring}.

\subsection{Model and observables}
For each network realization, Boolean functions are obtained operationally by selecting $k$ input nodes, iterating over all $2^k$ binary input configurations, and running a cascade for each input configuration.  The resulting activation pattern of any non-input node defines a $k$-ary Boolean function mapping inputs to outputs.

We studied linear-threshold cascades on Erd\H{o}s--R\'enyi networks with $N=10^4$, homogeneous threshold $\phi = 0.1$, and $k = 5$ externally clamped input nodes, averaging functional and information-theoretic observables over independent realizations as a function of mean degree $z$.  Across simulations, multiple measures of structure, function, and information processing exhibit pronounced changes at the structural percolation threshold $z_c \approx 1$ for fixed 
$k$ and fixed threshold parameters [Fig.~\ref{fig:functional_percolation}].

At $z_c$, the emergence of the giant connected component coincides with a sharp onset of typical decision-tree complexity (DTC), marking a sudden expansion in the complexity of realizable response functions once global connectivity is established.  The trial-to-trial variance of DTC grows markedly as the system enters the critical regime, consistent with a susceptibility-like enhancement of fluctuations near a phase transition.  Simultaneously, the mean number of unique Boolean functions realized by the network increases rapidly near $z_c$, indicating an abrupt growth in accessible functional diversity at the transition.

Information-theoretic measures reveal closely related but more nuanced behavior.  The output entropy of non-input nodes rises sharply at $z_c$ and reaches a pronounced maximum slightly above the structural percolation point, indicating a regime of maximally diverse and balanced responses under uniform input stimulation.  Beyond this point, output entropy decreases modestly as increasing connectivity constrains responses through stronger collective coupling.  Transfer entropy (TE) also rises sharply at $z_c$, signaling the onset of globally transmissible dynamical influence, and continues to increase with connectivity as directed information flow becomes progressively stronger in the supercritical regime.

Together, these results demonstrate that structural percolation induces a qualitative transition in both functional capacity and information processing.  Functional diversity and response expressivity are maximized near criticality, while sustained increases in connectivity favor increasingly directed information flow, delineating distinct functional regimes separated by the percolation threshold.

\begin{figure}[h!]
    \centering
    \includegraphics[width=\linewidth]{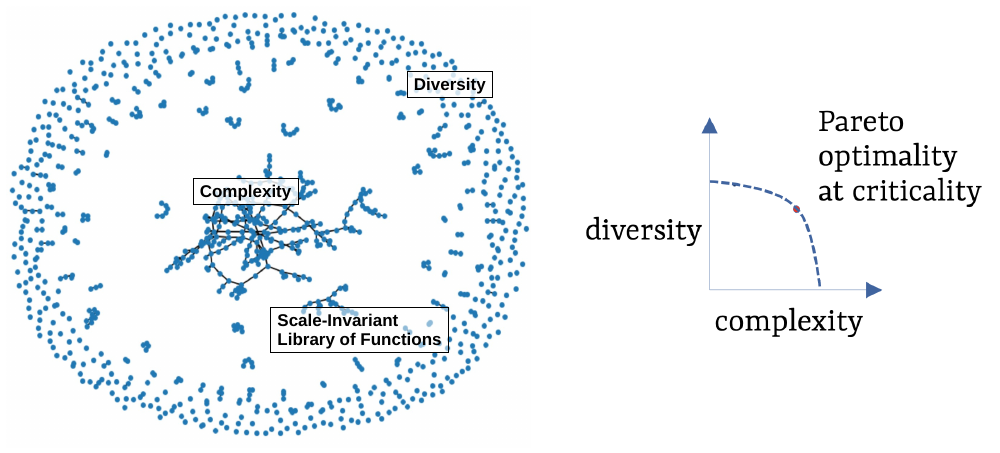}
    \caption{\textbf{Pareto optimality---Complexity versus diversity.}  
[Left] The formation of the giant connected component (GCC) at criticality allows the system to perform complex information processing and compute high-DTC functions.  
At the same time, the scale-invariant distribution of finite components at percolation---previously shown to give rise to small functional motifs---provides a natural source of computational diversity.  
[Right] This competition gives rise to an apparent Pareto optimality between functional complexity and diversity.}

    \label{fig:Pareto_optimality}
    \par\smallskip\noindent{\footnotesize\textit{Alt text:} Two panels. Left panel: a network at percolation criticality, showing one central connected giant component (labeled ``Complexity'') surrounded by many small scattered finite clusters (labeled ``Diversity'' and ``Scale-Invariant Library of Functions''). Right panel: a schematic convex trade-off curve between complexity (horizontal axis) and diversity (vertical axis), with a marked point on the curve indicating Pareto optimality at criticality.}
\end{figure}

At criticality, the emergence of the giant connected component is accompanied by a scale-invariant distribution of finite components, which have been shown previously to act as small functional motifs in cascade dynamics \cite{Wilkerson2022}.
Due to the simultaneous growth of both DTC and functional diversity near criticality, we observe that this point appears to represent a compromise at \textit{Pareto optimality between complexity and diversity of information processing} \cite{Langton1990EdgeOfChaos} [Fig.~\ref{fig:Pareto_optimality}].  Exceeding the critical point suppresses exploratory capacity through over-constrained collective dynamics, while limited connectivity below criticality leads to fragmentation and an inability to realize complex functions.  Thus, in the context of the overwhelming size of the Boolean function space, a system optimizes its exploration of diverse, complex functions by operating near criticality.

These considerations motivate the interpretation of criticality as a pivot point between two distinct functional regimes: one supporting the \textit{execution} of complex, integrative computations, and another enabling efficient \textit{exploration} of function space.  From this perspective, criticality may be viewed as a natural operating point for balancing exploration and exploitation, a distinction that may be framed phenomenologically in terms of reinforcement learning in adaptive or neural systems \cite{Fisher1967,Mora2011,KinouchiCopelli2006}.

\section{Discussion and Conclusion}

Our focus here has been on the physical precursors of information processing, framing them in terms of the most basic and universal physical processes.  While these may seem to underpin eventual Turing complete computation, or statistical models of learning, meriting further exploration in those contexts, we do not yet make those claims here.
Rather, we begin from first principles of universal characteristics in physical systems to understand what emerges.
We observe that cascades in random networks provide a generic information processing substrate, representing dynamics in thresholded, information propagating systems, underlying and independent of specifics found in biological, artificial, or social systems.

From our results [Figure \ref{fig:functional_percolation}], we observe an agreement between functional and information-theoretic criticality.   Transfer entropy can provide a complementary indicator of directed dynamical influence \cite{schreiber2000measuring}.  This gives support to criticality of information processing beyond Boolean functionality.
It also offers a structural-functional mechanism underlying the neural \textit{criticality hypothesis}, that optimal information processing happens near criticality, giving us guidance of this behavior from first principles \cite{BeggsPlenz2003,Shew2009DynamicRange,KinouchiCopelli2006}.
In the context of the scale mismatch between configuration space and function space discussed above, functional percolation thus appears as a natural way to reconcile limited structural resources with rich information-processing capabilities.

Here we have observed how structural constraints can shape information states.  Also, our results for functional diversity [Figure \ref{fig:functional_percolation}] suggest that, near criticality, the system can access a wide range of functions without needing extensive physical rewiring or many distinct configurations.  Thus, remaining near criticality minimizes the structural 'work' needed to explore information-processing states, because function-space exploration is maximized per unit structural change, and fewer physical modifications are needed to reach many computational states.  Due to the universal nature of the scale mismatch outlined above, this suggests that networks must use every possible tool at their disposal to efficiently explore the function space.

We note that interaction topology and information propagation in networks allow us to begin to connect Boltzmann and Shannon entropies in a direct way, grounded in explicit causal interactions rather than ensemble-level statistical assumptions \cite{sethna2001crackling,Jaynes1957}.
At a deeper level, the simultaneous increase of functional complexity, entropy, and information flow at criticality reflects the removal of structural constraints at the percolation transition, which dramatically expands the accessible state space of the system; while this invites a thermodynamic interpretation in terms of effective entropy under coarse-graining, we do not attempt a formal entropy-balance or energetic treatment here.  While we are not here making exact formal statements about Landauer bit-erasure or thermodynamics \cite{Landauer1961Irreversibility}, it is very compelling to undertake more formal exploration.

We note here that the structure--function scale-mismatch argument above motivates additional search mechanisms by networks beyond static connectivity alone.  Criticality itself is likely insufficient to span the exponential scale gap.  Dynamics, for example in oscillatory systems, provide a plausible way for systems to reduce dependence on fixed physical connectivity by creating transient virtual computational circuits overlaying the physical topology on short timescales.  
Kuramoto oscillators, for example, can be said to be `connected'---transmitting information---when nearly in phase, but effectively `disconnected' when out of phase \cite{Kuramoto1975,schreiber2000measuring}, enabling rapid reconfiguration of functional interactions.  One may speculate that such virtual circuit rewiring allows neural systems to explore functional responses using spiking and phase dynamics orders of magnitude faster than physical growth or rewiring \cite{Fries2005-coherence,Varela2001-brainweb,DecoKringelbach2016-metastability}.  
From a similar perspective, inhibitory neurons may play a complementary role by dynamically reducing effective connectivity, potentially stabilizing networks near criticality in a form of self-organizing criticality (SOC) that facilitates efficient exploration of function space \cite{bak1988self,levina2007dynamical}.  
While these behaviors have been empirically observed and studied, it is noteworthy that they align closely with mechanisms suggested independently by first-principles arguments and combinatorial constraints.

Due to the compelling fundamental nature of the scale mismatch outlined above, one may expect that analogous optimizing mechanisms arise in other physical and biological systems.

Because our analysis begins from minimal physical and abstract mechanisms---local state encoding, thresholds, propagating interactions, and structural connectivity---it offers a substrate-independent perspective on how systems can explore and realize complex input--output transformations. In this sense, functional percolation may contribute to a deeper theoretical foundation for learning and information processing in physical and artificial systems.

Finally, due to the basic and system-independent nature of the dynamics outlined here, we propose that \textit{functional percolation is a candidate unifying physical mechanism}.  There is a coincidence of structural, functional, and informational phase transitions. 
While percolation theory is formally defined in the $N \to \infty$ limit, the sharp transitions observed here at large system size across multiple functional and information-theoretic order parameters are consistent with the existence of a well-defined functional percolation transition in the thermodynamic limit.
In this sense, functional percolation is not only a useful operating regime, but appears to be a generic consequence of the structural and combinatorial constraints imposed on information processing in extended physical systems.  This motivates extensive further investigation and formalization of the topics discussed here.

\section*{Funding}
This work received no specific funding.

\section*{Competing interests}
The author declares no competing interests.

\providecommand{\newblock}{}

\end{document}